\documentclass[paper]{JHEP}   
\usepackage{amsmath}
\usepackage{amssymb}
\usepackage{cite}

\usepackage{epsfig}

\psfull

\newcommand{\ie}{\emph{i.e.}\ }
\newcommand{\eg}{\emph{e.g.}\ }
\newcommand{\cnf}{\emph{cf.}\ }

\def\order#1{{\cal{O}}\left(#1\right)}

\def\conf{\delta}

\def\ktilde{{\tilde k}}

\def\qbar{{\bar q}}

\def\cF{{\cal{F}}}    

\def\MSbar{\overline{\mbox{\scriptsize MS}}}

\def\CF{C_F}

\def\CA{C_A}

\def\nf{n_{\!f}}

\def\as{\alpha_{{\textsc{s}}}}

\def\cO#1{{\cal{O}}\left(#1\right)}

\def\ee{e^+e^-}

\title{Semi-numerical resummation \\ of event shapes
\thanks{Research supported in part by the EU Fourth Framework 
Programme `Training and Mobility of Researchers', Network `Quantum
Chromodynamics and the Deep Structure of Elementary Particles',
contract FMRX-CT98-0194 (DG 12-MIHT).}}

\author{
Andrea Banfi,\\ 
Dipartimento di Fisica, Universit{\`a} di Milano--Bicocca
and INFN, Sezione di Milano, Italy} 

\author{
 Gavin P.~Salam \\
       LPTHE, Universit\'es P. \& M. Curie (Paris VI) et D. Diderot
       (Paris VII), Paris, France.\\
       CERN, TH Division, 1211 Geneva 23, Switzerland.
}
\author{
Giulia Zanderighi \\ 
       Department of Physics, University of Durham, Durham DH1 3LE, England.
} 

\abstract{
  For many event-shape observables, the most difficult part of a
  resummation in the Born limit is the analytical treatment of the
  observable's dependence on multiple emissions, which is required at
  single logarithmic accuracy. We present a general numerical method,
  suitable 
  for a large class of event shapes, which allows the resummation
  specifically of these single logarithms. It is applied to the case
  of the thrust major and the oblateness, which have so far defied
  analytical resummation and to the two-jet rate in the Durham
  algorithm, for which only a subset of the single logs had up to now
  been calculated.
}

\preprint{
  Bicocca--FT--01--28\\
  CERN--TH/2001--352\\
  DCTP--01--122\\
  IPPP--01--61\\
  LPTHE--01--77\\
  hep-ph/0112156 \\
  December 2001}

\begin{document}

\section{Introduction}

In QCD, for many observables of interest, the only calculational
approach that is currently available is a perturbative expansion in
powers of the strong coupling $\as$.  Quite often the coefficients of
this expansion are well behaved and reasonable accuracy can be
obtained given the first two or three terms of the series.

However there exist whole classes of observables for which the
coefficients grow rapidly order by order. Typically this happens in
problems involving two or more scales, in which case the coefficient
of $\as^n$ may involve up to $2n$ factors of the logarithm, $L$, of
the ratio of scales. These logarithms are generally associated with
soft and collinear regions of phase space.  When $L$ is
large it compensates the smallness of $\as$ and reliable predictions
can be obtained only by including an infinite number of terms. While
significant technical difficulties prevent an exact determination of
high-order terms, it is sometimes
possible to calculate the coefficient of their dominant (and
sub-dominant) logarithmic parts, order by order, and \emph{resum} this
subset of terms. In many cases this is sufficient to obtain a reliable
prediction.

One class of observables that has been extensively studied in this
context is that of event
shapes~\cite{CTTW,CTW,Cpar,NewBroad,DIStauz,DISBroad,rhoLight,KoutLong,Dpar,KoutHH,KoutDIS,1jet}
and jet rates~\cite{JetRates,DissSchmell,4JetRates}. These measure
geometrical properties of the hadronic energy flow in the final state
of a given reaction, \eg in $\ee \to \mathrm{hadrons}$, and are widely
used in tests of QCD, the measurement of $\as$, and studies of the
interface between perturbative and non-perturbative physics. When the
value $V$ of the observable is small, the perturbative expansion of
its distribution is dominated by terms $(\as^n \ln^{2n-1} V)/V$ and a
resummation is necessary.

The determination of the order $\as$ and $\as^2$ contributions to such
observables is usually fairly straightforward.
One writes a computer subroutine which calculates
the observable for an arbitrary set of 4-momenta and links it with a
program which codes the fixed-order real and virtual matrix elements
for a given process. The `hard' part of the problem is to determine
the matrix elements and code them into the program so as to obtain a
numerically efficient cancellation of real and virtual contributions.
But this part of the problem is common to all infrared and collinear
safe observables and is `ready-solved' in the form of a range of
publicly-available programs, such as EVENT2 and DISENT
\cite{CSDipole}, and also many others, see for example
\cite{JETRAD,Disaster,DebrecenFamily,EERAD2,MenloParc,Mercutio}.

On the other hand the determination of resummed distributions of event
shape observables has up to now proved more labour-intensive. This is
despite the fact that many aspects of the calculation are actually
fairly similar from one observable to another --- for example one can
generally make the same set of approximations concerning the
multi-particle matrix-element, writing it as a product of matrix
elements for independent emissions. However it is also necessary to
relate the value of the observable to some condition on the ensemble
of emissions. For a resummation to be feasible analytically, one must
express this condition as a product of factors each involving a
single emission. It is this part that requires a separate analytical
calculation for each observable.

To explain in more detail what we mean, it is useful to consider the thrust
$T$ in $\ee$ (defined in appendix~\ref{sec:ObsDef}).\footnote{The
  thrust is slightly unusual in that the two-jet region (where the
  resummation is needed) corresponds to $T\to1$, but this is just a
  matter of definition.} In the two-jet region its value is given by a
sum over contributions from individual emissions:
\begin{equation}
  \label{eq:ThrustMulti}
  T = 1 - \sum_i \frac{k_{ti}}{Q} e^{-|\eta_i|}\,.
\end{equation}
In performing the resummation one determines the cross section for
$1-T$ to be smaller than some value, say $1-T < \tau$. Using an
integral transform this condition can be written as
\begin{equation}
  \label{eq:Factorise}
  \Theta(\tau - (1 - T)) = \int \frac{d\nu}{2\pi i \nu} \, e^{ \nu \tau }
  \prod_i \exp \left(-\nu \frac{k_{ti}}{Q} e^{-|\eta_i|} \right)\,,
\end{equation}
which is a product of factors for individual emissions, just as
required. 

The thrust is a particularly simple example, because it is easy both
to determine its dependence on multiple emissions,
eq.~\eqref{eq:ThrustMulti}, and to factorise this dependence,
eq.~\eqref{eq:Factorise}. Other variables are less trivial. For
example for the jet-broadenings the determination of the dependence on
multiple emissions involves subtle issues of hard-parton recoil, and
the factorisation requires at least two integral
transforms~\cite{NewBroad}. Some multi-jet event shapes, such as the
thrust minor in the three-jet limit involve as many as five additional
Fourier transforms~\cite{KoutLong}!  This makes the resummation of
event shapes quite tedious (and sometimes even error
prone). Furthermore for some variables, for example the
thrust major and oblateness in $\ee$, it is not even possible to write
down a closed expression for the dependence of the observable on a
general multi-particle configuration, rendering an analytical
resummation unfeasible.

So in this article we take a step towards the development of a
systematic semi-numerical approach for calculating resummed
distributions for a large class of observables, essentially those for
which double logarithms exponentiate. The
probability, $\Sigma(v)$, of a suitable  observable having a value smaller
than $v$ is given by an expression of the form
\begin{equation}
  \label{eq:SigmaGen}
  \Sigma(v) = \exp\left[ L g_1(\as L)  + g_2 (\as L) + \cdots
  \right]\,,
  \qquad L = \ln 1/v\>.
\end{equation}
The terms in $L g_1(\as L)$ are known as leading, or double
logarithms (LL, DL), and start at $\order{\as L^2}$.  The terms in
$g_2(\as L)$ are known as next-to-leading, or single logarithms
(NLL, SL), and start at $\order{\as L}$. The double logarithms arise from
a veto of a soft and collinear portion of phase space and are
straightforward to calculate. The single logarithms arise from a
variety of sources.  For global variables \cite{1jet} most of the
sources (\eg from the running coupling, hard collinear emission,
angles between hard partons) can be understood by considering the
phase space for just a single emission.  What remain are single
logarithms associated with the observable's dependence on multiple
emissions. It is this contribution that we address here.

Our method involves first considering a `simple' reference variable,
in which the dependence on multiple emissions is trivial, leading to a
straightforward analytical resummation. One then numerically relates
the resummation of the simple variable to that of a more complicated
variable of our choosing. For this to work the two variables must have
the same structure of double logs.

So in section~\ref{sec:proc} we show how to relate a pair of variables
with the same double logs, and give a procedure for designing a
suitable `simple' reference variable. We then show how this can be
implemented in a Monte Carlo algorithm for calculating the
single-logarithmic function relating the resummation of the two
variables.

In section~\ref{sec:knowneg} we show that this method reproduces known
results, with examples from $\ee \to 2$~jets and $\ee \to 3$~jets.

Finally in section~\ref{sec:neweg} we present new results for some
widely studied variables in $\ee \to 2$~jets: for the thrust major and
the oblateness for which no resummation had ever been performed up to
now (and which are probably beyond analytical treatment) and for the
Durham three-jet resolution parameter for which resummed results
existed but not to full NLL order~\cite{JetRates,DissSchmell}.

\section{General procedure}
\label{sec:proc}

This section sets out the principal elements of our approach. It is
divided into several parts. First we see how, at NLL accuracy, to
relate the distributions of two variables, starting from knowledge of
the probability of distribution of one variable given the value of the
other.  

We then see how to design a `simple' reference variable for which the
analytical resummation is trivial and examine how to calculate the
probability distribution of a more complicated variable given the
value of this simple one.

The section finishes with the consideration of the subtle but
practically important issues of how to ensure that our final numerical
calculation is free of contamination from next-to-next-to-leading
logarithmic (NNLL) corrections, and how to determine its expansion to
second order in $\as$ (needed for the subtraction of doubly counted
terms when matching to fixed order results).

\subsection{Relating the resummations of two observables}

Suppose we have a `complicated' observable $V$ and a `simple'
observable $V_s$ with the same all-orders double logarithmic structure
(terms $\as^n L^{n+1}$ in the exponent). We introduce the probability
distribution $P(v|v_s)$ for the value $v$ of the complicated
observable, given a value $v_s$ for the simple one.

Writing the distribution of the simple observable as $D_s(v_s)/v_s$,
one obtains the distribution of the more complex observable $D(v)/v$
through the following convolution:
\begin{equation}
  D(v) = v \int \frac{dv_s}{v_s} \, D_s(v_s) \,P(v | v_s)\>.
\end{equation}
If $P(v|v_s)$ is dominated by the region $v \sim v_s$ (this will
follow naturally from the two variables having the same double-log
structure) then one may expand the distribution $D_s(v_s)$ using
eq.~\eqref{eq:SigmaGen},
\begin{equation}
  D_s(v_s) = D_s(v)\, \exp \left(- R'\ln \frac{v}{v_s} + \order{\as^n
      L^{n-1} \ln^2 \frac{v}{v_s}} \right)\>,
\end{equation}
where $R'$, which is related to the (differential) phase space for
emissions, is
defined as
\begin{equation}
  R' = - \frac{d\ln \Sigma}{dL}\>.
\end{equation}
Neglecting NNLL terms, $\as^n L^{n-1}$, this gives
\begin{equation}
  \label{eq:DvDsv}
  D(v) = v D_s(v) \int \frac{dv_s}{v_s} \, e^{-R' \ln v/v_s} \,P(v |
  v_s)\,. 
\end{equation}
As we shall see shortly, for suitable observables (as defined in
appendix~\ref{sec:applicability}), to NLL accuracy $v P(v|v_s)$
depends only on the ratio $v/v_s$ and on the phase space for
emissions (through $R'$):
\begin{equation}
  P(v|v_s) = \frac{1}{v}\, p\!\left(\frac{v}{v_s}, R'\right) \left(1
    + \order{\as^n
      L^{n-1} \ln \frac{v}{v_s}}\right ),
\end{equation}
where we have introduced a rescaled probability $p(x,R')$ for the ratio
$v/v_s$ to be to equal to $x$ (with measure $dx/x$). As long as $v$
and $v_s$ are of the same order, the ambiguity of whether $R'$ should
be evaluated at $v$ or $v_s$ corresponds to a subleading effect since
$R'(v) = R'(v_s) + \order{\as^n L^{n-1} \ln v/v_s}$.
This allows us to write
\begin{equation}
  \label{eq:Drelation}
  D(v) = D_s(v) \cF(R'),
\end{equation}
where
\begin{equation}
  \label{eq:FDefn}
\cF(R')  = \int \frac{dx}{x} \, e^{-R' \ln x} 
\,p\left(x, R'\right)\,.
\end{equation}
A relation analogous to eq.~\eqref{eq:Drelation} holds also for the
integrated distributions 
\begin{equation}
  \label{eq:Sigmarelation}
  \Sigma(v) = \Sigma_s(v) \cF(R'),
\end{equation}
as can be seen by differentiating and noting that extra term relative
to \eqref{eq:Drelation} is subleading:
\begin{equation}
  \frac{d}{dL}\Sigma(v) =  \left( 
    \cF(R') + \frac{R'' \frac{d \cF}{dR'}}{\frac{d}{dL} \ln \Sigma_s}
    \right)
    \frac{d}{dL }\Sigma_s(v)\,.
\end{equation}

\subsection{The `simple' observable}

So far the simple observable has been left as a fairly vague concept.
Here we outline a concrete procedure for constructing it.

Given a Born configuration consisting of hard momenta $\{p_\mathrm{Born}
\}$ together with an arbitrary set of soft and collinear emissions
$k_1, \ldots, k_n$, a general way of defining the simple observable is as
follows:
%
%
\begin{multline}
  \label{eq:VsDef}
  V_s(\{p_\mathrm{Born}\}, k_1, k_2, \ldots, k_n) =\\ 
  \max\left[ V(\{p_\mathrm{Born}\}, k_1), V(\{p_\mathrm{Born}\}, k_2), \ldots,
    V(\{p_\mathrm{Born}\}, k_n) \right]\,.
\end{multline}
This definition is unsafe with respect to secondary collinear
branching of the soft and collinear emissions. However for the
`suitable' class of event shapes under consideration, secondary
collinear branching is irrelevant except insofar as it is responsible
for determining the scale of the coupling \cite{CTTW} --- accordingly
to calculate the resummation of the simple observable and $\cF(R')$ we
consider only independent emissions from the Born configuration and
account for collinear splitting by directly setting the
scale of the coupling.\footnote{One could perhaps devise a
  collinear-safe definition which for our purposes is equivalent to
  the above one. This might involve the application of a
  clustering algorithm to emissions before using
  eq.~\eqref{eq:VsDef}.}

Having dealt with this issue, the resummation for $V_s$ is then
simple, because the condition for the value of the observable to be
smaller than $v_s$ factorises straightforwardly,
\begin{equation}
  \label{eq:FactoredVs}
  \Theta(v_s - V_s({k_1,k_2,\ldots,k_n})) 
  \equiv  \prod_{i=1}^n \Theta(v_s - V(k_i))\,,
\end{equation}
where for brevity we have dropped the Born momenta from the arguments
of $V$.  

To illustrate this we shall consider the class of global observables
in $\ee \to 2$~jets. Emissions can be taken to be independent, as
shown in \cite{CTW,NewBroad}, giving
the following expression for their probability distribution:
\begin{equation}
  \label{eq:IndepEmsnProb}
  dP({k_1,\ldots,k_n}) = e^{-R_{\epsilon}} 
  \frac{1}{n!} \prod_{i=1}^n
  d\eta_i \frac{d{k}^2_{ti}}{k_{ti}^2} \frac{d\phi_i}{2\pi} \frac{\CF
    \as(k_{ti})}{\pi} \,
  \Theta\left(\ln \frac{Q}{k_{ti}} -\frac34 - |\eta_i|\right)\,,
\end{equation}
where $k_{ti}$, $\eta_i$ and $\phi_i$ are respectively the transverse
momentum, the rapidity and the azimuthal angle of $k_i$ with respect to
the $q\qbar$ axis.  Virtual corrections are given by the
factor $e^{-R_\epsilon}$, with
\begin{equation}
  R_{\epsilon} = \int d\eta \frac{d{k^2_t}}{k_{t}^2} \frac{d\phi}{2\pi}
  \frac{\CF \as(k_t)}{\pi} \,
  \Theta\left(\ln \frac{Q}{k_t} -\frac34 - |\eta|\right),
\end{equation}
where the $\epsilon$ indicates that some regularisation must be
applied to both the real and virtual parts. In the above formulae, the
coupling is defined in the gluon Bremsstrahlung scheme \cite{CMW} and
the $-3/4$ 
in the limit on the rapidity accounts for the hard part of the
$P_{qq}$ collinear splitting function (see for example
\cite{NewBroad}). Combining these expressions 
with eq.~\eqref{eq:FactoredVs} gives
\begin{equation}
\label{eq:Sigmas}
  \Sigma_s(v_s) = \exp\left( - R_s(v_s) \right)\,,
\end{equation}
with
\begin{equation}
\label{eq:Rs}
  R_s(v_s) = 
  \CF \int^{Q^2} \frac{dk_t^2}{k_t^2} \frac{d\phi}{2\pi}
  \frac{\as(k_t)}{\pi} 
    \int^{\ln \frac{Q}{k_t} - \frac34}_{-\ln \frac{Q}{k_t} + \frac34}d\eta \,
    \Theta(V(k)- v_s)\,.
\end{equation}

In processes with three hard partons the resummation of the simple variable
also proves straightforward, with $R_{s}$ just involving a sum over
dipoles \cite{KoutLong,Dpar,KoutHH,KoutDIS}. On the other hand,
with 4 or more hard partons subtleties will arise from the large angle
region as can be seen from \cite{StermanEtAl}.\footnote{Though it
  remains to be seen whether the extensive technology developed in
  \cite{StermanEtAl} accounts for the non-global effects that are to
  be expected for the observables that are considered there.}

\subsection{The probability $\boldsymbol{P(v|v_s)}$}

One of the advantages of the above choice for the `simple' observable
is that it is easy to design a Monte Carlo algorithm to construct, for
a given $R'$, an ensemble of configurations which all have the same
value of $v_s$ and known (equal) weights.

Fixing $v_s$ implies taking only those configurations where one of the
emissions, $j$, satisfies $V(k_j) = v_s$, and all others $i\ne j$
satisfy $V(k_i) < v_s$. Using eq.~\eqref{eq:IndepEmsnProb},
renumbering emissions such that $j$ is always $1$, and dividing by the
total probability of the simple variable having value $v_s$, we obtain
the following expression for the probability distribution of
configurations given $v_s$:
\begin{multline}
  \label{eq:CondProbKs}
  dP({k_1,\ldots,k_n} | v_s) = \frac{v_s}{R_s'(v_s)}
  \frac{e^{-R_{\epsilon} + R_s(v_s)}}{(n-1)!} \times \delta(v_s - V(k_1))
  \prod_{i=2}^n \Theta(v_s - V(k_i))\\
\left[ \prod_{i=1}^n
    d\eta_i \frac{d{k^2_{ti}}}{k_{ti}^2} \frac{d\phi_i}{2\pi} \frac{\CF
      \as(k_{ti})}{\pi} \Theta\left(\ln \frac{Q}{k_{ti}} -\frac34 -
      |\eta_i|\right) \right]\,,
\end{multline}
where
\begin{equation}
  R'_s(v_s) = - v_s \frac{d R_s}{dv_s} = R'(v_s) + \order{\as^{n} L^{n-1}}\,.
\end{equation}
It is now straightforward to construct $P(v|v_s)$:
\begin{equation}
  \label{eq:PvGivenVs}
  P(v|v_s) = \sum_{n=1}^\infty \int dP({k_1,\ldots,k_n} | v_s)\,
  \delta(V({k_1,\ldots,k_n}) - v)\,.
\end{equation}

\subsection{Eliminating subleading effects}
\label{sec:nosubleading}

As they stand, the above expressions contain not just single-log terms
but also many (sometimes spurious) subleading contributions. For
example while in eq.~\eqref{eq:Rs} the `$3/4$' is necessary in order
to obtain the correct single (NL) logs, in eq.~\eqref{eq:CondProbKs}
it leads only to NNLL terms. These subleading terms
are awkward for a variety of reasons. Firstly there is currently no
way of guaranteeing their correctness. 
Secondly when matching to fixed-order predictions they lead to
some technical difficulties because one needs a good knowledge of
their expansion to $\order{\as^2}$. Finally one notes that the
expressions of the previous section contain integrals over $\as(k_t)$
down into the infrared. The Landau pole in $\as$ (or whatever other
structure occurs in higher orders) will then introduce a power
correction ambiguity in one's answer, requiring that one make an
arbitrary non-perturbative `choice' concerning its treatment.
Accordingly it has become standard to give predictions containing
nothing but leading and NL logs (though we note the alternative
philosophy advocated in \cite{GardiGrunbergRathsman}).

Truncating the expressions for $R_s$ and $R'$ at NLL order is easy,
since they are computed analytically. However ensuring that $\cF$ is
purely a function of $R'$, required in order for it to have just NLL
terms, turns out to be the most subtle part of our semi-numerical
resummation approach.  We present two ways of doing it.

\paragraph{Method 1 (general).} One way of eliminating subleading
effects is by 
numerically taking the limit of $\as \to 0$, while keeping $\as \ln
1/v$ fixed (\ie fixed $R'$).  While conceptually straightforward
and very general, such an approach turns out to be numerically
awkward: the NNLL contamination in one's answer is of order $\as$, and
for fixed $R'$, $\ln 1/v$ scales roughly as $1/\as$. 
So if for
argument's sake we want to eliminate subleading effects to within a
percent, then we need to consider values of $v$ of the order of
$10^{-100}$ (since we are dealing with orders of magnitude of the
logarithm, we 
ignore the difference between $e^{-100}$ and $10^{-100}$).
Unfortunately standard algorithms for calculating event 
shapes generally work only for $v \gg \varepsilon^p$ where
$\varepsilon$ is the relative floating precision (usually $10^{-15}$ in double
precision) and $p$ is some algorithm-dependent power of order $1$.

So in order for this approach to work we need to find an algorithm for
calculating the observable that is free of sensitivity to the limited
accuracy of the floating-point arithmetic.\footnote{An alternative
  approach, whose feasibility we have not investigated in detail, would
  be to use arbitrary precision arithmetic. This would entail a
  significant penalty in terms of computing time needed, but might
  nevertheless be an approach worthy
  of investigation.} %
Additionally the algorithm
should work with a representation of 4-momenta that does not suffer
from the limit on the smallest representable floating-point
number (in standard double precision the smallest number that can be
represented is of the order of $10^{-308}$).
Both these issues restrict the generality of the
approach, and imply a certain amount of analysis of one's observable
so as to understand the elements fundamental to its determination in
terms of a set of soft and collinear momenta. This will be illustrated
in detail in section~\ref{sec:Durham} when we come to discuss the
2-jet rate for the Durham algorithm.

\paragraph{Method 2 (event-shape specific).} Fortunately there exists
a much simpler approach 
which works in almost all cases. Event shapes (but not jet rates) have
the following useful property: given an ensemble of soft and collinear
momenta $\{k_1,\ldots, k_n\}$, if we vary the $\{k_i\}$ such that the
$\{V(k_i)\}$ are kept constant and the $\{k_i\}$ all remain collinear
to the legs to which they were originally collinear, then $V(\{k_i\})$
also remains constant. In eq.~\eqref{eq:PvGivenVs} this allows us to
carry out the integration over rapidity analytically and replace it
with a sum over the different hard legs to which an emission may be
collinear (in calculating $\cF$ the large-angle region gives
only NNLL terms).

So we divide the phase space $R'$ into contributions $R_\ell'$ coming from $N$
different hard legs
\begin{equation}
\label{eq:Rpell}
  R'(v) = \sum_{\ell=1}^N R'_\ell (v)\,,
\end{equation}
and write
\begin{multline}
  dP(\ktilde_1, \ldots, \ktilde_n | v_s)
  = \frac{v_s}{R_s'(v_s)}
  \frac{e^{-R_{\epsilon} + R_s(v_s)}}{(n-1)!}
  \left[\prod_{i=1}^n \frac{dv_i}{v_i} \frac{d\phi_i}{2\pi}
    R'_{\ell_i} 
    (v_i)\right] 
  \\ \times
  \delta(v_s - v_1)
  \prod_{i=2}^n \Theta(v_s - v_i)\,, 
\end{multline}
where the use of $\ktilde_i$ rather than $k_i$ denotes the fact that
emissions are characterised by the leg $\ell_i$ to which they are
collinear, their azimuthal angle $\phi_i$ and the value of $v_i$. The
transverse momentum and rapidity are no longer individually defined
--- they have been integrated over while keeping $v_i$ constant.

To eliminate subleading logarithms, we now exploit the infrared safety
of event shapes, which implies that the emissions that contribute
significantly to $V(\{k_i\})$ are those for which $v_i \sim v_1$.
Accordingly we can throw away logs of $v_1/v_i$ (both in the real and
virtual parts) to get
\begin{equation}
  \label{eq:finalPks}
  dP(\ktilde_1, \ldots, \ktilde_n | v_s)
  = \frac{v_s}{R'}
  \frac{e^{-R' \int^{v_s} \frac{dv}{v} }}{(n-1)!}
  \left[\prod_{i=1}^n \frac{dv_i}{v_i} \frac{d\phi_i}{2\pi}
    R'_{\ell_i} 
    \right] 
  \delta(v_s - v_1)
  \prod_{i=2}^n \Theta(v_s - v_i)\,, 
\end{equation}
where the $R'_\ell$ are all evaluated at $v_s$. By
truncating their analytical expressions at SL level, we obtain an
answer for $dP(\ldots|v_s)$ that has only single logarithmic
dependence on $v_s$. Finally to get the equivalent of
eq.~\eqref{eq:PvGivenVs} we have to remember not only to integrate
over the $\phi_i$ and $v_i$ but also to sum over all the $\ell_i$:
\begin{equation}
 \label{eq:finalPkssum}
  p\left(\frac{v}{v_s},R'\right) = v
  \sum_{n=1}^\infty  \int
  \sum_{\ell_1\!,\ldots,\,\ell_n}
  dP({\ktilde_1,\ldots,\ktilde_n} | v_s)\, 
  \delta(V({\ktilde_1,\ldots,\ktilde_n}) - v)\,.
\end{equation}
We have written the answer for $p(v/v_s,R')$ rather than for
$P(v|v_s)$ so as to emphasise that we are free to choose $v_s$ as we
like, and that the result is a function only of $v/v_s$ and $R'$, free
of subleading logarithms (we recall that at NLL order, $R'$ itself
depends only on $\as L$). In practice there is still some
contamination from power suppressed contributions associated with the
observable itself, terms of relative order $v_s^q$, with $q$ some
observable dependent power of order $1$. However it is usually
perfectly feasible to choose values of $v_s$ such that $v_s^q \ll 1$,
while keeping the relative rounding errors associated with the
numerical calculation of the observable, $\varepsilon^p/v_s$, also
negligible.

For the calculations whose results are given below, the following
Monte Carlo algorithm has been used to generate configurations
according to the probability distribution in eq.~\eqref{eq:finalPks}.
One starts with $i=1$ and $y_1 = 1$, then:
\begin{itemize} \label{page:applic}
\item[1.] One chooses a leg $\ell_i$ randomly, the probability of leg
  $\ell$ being $R'_\ell / R'$.
\item[2.] One chooses uniformly a random azimuthal angle, $\phi_i$,
  with respect to the leg.
\item[3.] One then chooses the transverse momentum $k_{ti}$ and
  rapidity $\eta_i$ with respect to the leg, such that the emission is
  soft, collinear and satisfies $V(k_i) = y_i v_s$. Within these
  constraints one has total freedom in one's choice of transverse
  momentum and rapidity.
\item[4.] One chooses $\ln y_{i+1}$ such that $y_{i+1} < y_{i}$, with a
  random distribution proportional to $(y_{i+1}/y_i)^{R'}$ reflecting
  the fact that the phase space to produce an emission with $yv_s <
  V_{k} < (y + \delta y)v_s$ is $R' \delta y/y$.
\item[5.] If $y_{i+1} < \epsilon \ll 1$, with $\epsilon$ some arbitrary
  small cutoff, one stops. Otherwise one replaces $i\to i+1$ and goes back to
  step 1.
\end{itemize}


\subsection{The coefficient of $\boldsymbol{{R'}^2}$}

For the purposes of matching with fixed order calculations it is
necessary to have the expansion of $\cF(R')$ in powers of $R'$
\begin{equation}
  \cF(R') = 1 + \sum_{n=1}^\infty \cF_n {R'}^n
\end{equation}
(from which one can deduce the expansion in terms of powers of $\as
L$). Since matching is currently usually carried out to NL order one
needs the first two terms. As a consequence of our definition of the
`simple' observable, the first term is automatically zero. The second
term can be obtained by expanding eqs.~\eqref{eq:FDefn} and
\eqref{eq:finalPkssum} to give
\begin{equation}
  \cF_2 = -\sum_{\ell_1\!,\,\ell_2} \left(
  \lim_{\as L\to0} \frac{R'_{\ell_1}R'_{\ell_2}}{{R'}^2} \right)
  \int \frac{d\phi_1 d\phi_2}{(2\pi)^2}  \int^{v_s}
  \frac{dv_2}{v_2} \ln \frac{V(\ktilde_1,\ktilde_2)}{v_s}
  \,,\qquad V(\ktilde_1) = v_s\,.
\end{equation}
We specify the limit $R' \to 0$ for the ratio of $R_\ell'/R'$ because
for some more complicated observables, such as $\tau_{zE}$ in DIS
\cite{DIStauz}, $R_\ell'/R'$ is not a pure number but rather itself a
function of $R'$.  We also note that despite the appearance of $v_s$
in the integral, $\cF_2$ is of course independent of $v_s$.

Finally we point out that if necessary (\eg for NNLO matching) one can
in a similar way derive corresponding expressions for yet higher order
terms in the expansion of $\cF$.

\section{Reproducing known results}
\label{sec:knowneg}

An important test of our method is that it reproduces known analytical
results. This section shows results for a variety of previously
calculated observables in configurations with $2$ and $3$ hard jets.
For definitions of the observables, the reader is referred to
appendix~\ref{sec:ObsDef}.

\subsection{$\boldsymbol{\ee \to} \mbox{2 jets}$: 
thrust and broadenings \label{sec:oldeg}}

Here we recover the known
results for the thrust $T$ and the jet broadenings (total $B_T$ and
wide $B_W$).  In each case we have tested both methods for the
elimination of subleading logs (section~\ref{sec:nosubleading}) and
verified that they give identical results.

For the thrust distribution the `simple' observable is given by
\begin{equation}
\tau_s\equiv \max_i \left\{\frac{k_{ti}}{Q}e^{-|\eta_i|}\right\}\>,
\end{equation}
The lowest line of figure~\ref{fig:Fthrust} shows the comparison between the
function $\cF_{\tau}$ and the corresponding exact result~\cite{CTTW},
\begin{equation}
\cF_{\tau}(R')=\frac{e^{-\gamma_E R'}}{\Gamma(1+R')}\>.
\end{equation}
In the broadening case we start from
\begin{equation}
B_s\equiv \max_i \frac{|\vec{k}_{ti}|}{Q}\>.
\end{equation}
The Monte Carlo procedure gives $B_L$ and $B_R$ (see
\eqref{eq:BR-BL}) and then $B_T$ and $B_W$ (see \eqref{eq:BT-BW}). 
The comparison between numerical and analytical
resummation~\cite{NewBroad},
\begin{equation}
\begin{split}
\cF_{B_T}(R')&=\frac{e^{-\gamma_E R'}}{\Gamma(1+R')}
\left(\int_1^{\infty}\frac{dx}{x^2}
\left(\frac{1+x}{4}\right)^{-R'/2}\right)^{2}\,,\\
\cF_{B_W}(R')&=\frac{e^{-\gamma_E R'}}{\Gamma^2(1+R'/2)}
\left(\int_1^{\infty}\frac{dx}{x^2}
\left(\frac{1+x}{4}\right)^{-R'/2}\right)^{2}\>,
\end{split}
\end{equation}
is given by the top two lines of figure~\ref{fig:Fthrust}.

\EPSFIGURE[ht]{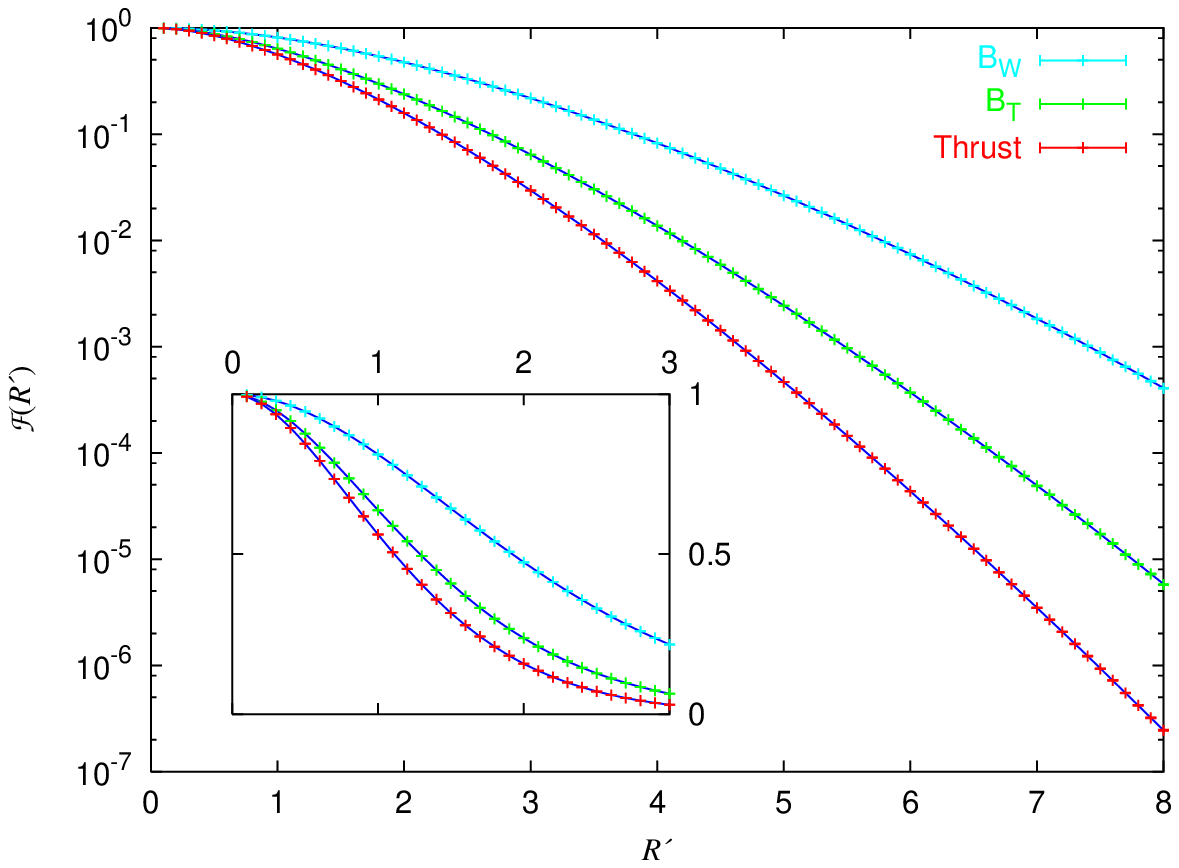,width=0.9\textwidth}{$\cF(R')$ for the thrust and the two broadenings. The lines are the analytical results, while the points are the numerical results.  \label{fig:Fthrust}}

\subsection{$\boldsymbol{\ee \to} \mbox{3 jets}$: 
thrust minor (a.k.a.\ $\boldsymbol{K_\mathrm{out}}$)}
The procedure can also be applied to multi-jet event shapes, such as
$T_m$ and $D$ parameter, which have been studied recently in the
near-to-planar three-jet region~\cite{KoutLong,Dpar}.  In particular, we
would like to compare the numerical results to the analytical
resummation of $T_m$ distribution, which is much more involved than
that of the $D$ parameter due to hard parton recoil effects.

In the near to planar three-jet region $T\sim T_M \gg T_m$, a
three-jet event consists of a hard quark-antiquark-gluon system
accompanied by soft secondary partons. We denote the hard partons by
$p_1,p_2$ and $p_3$ with $p_1^0>p_2^0>p_3^0$ and call $\conf=1,2,3$
the configuration in which the gluon momentum is $p_{\conf}$.  
As discussed in~\cite{KoutLong}, the function $\cF_{T_m}(R')$ depends
on the colour configuration of the hard underlying system.

The simple observable $T_{m,s}$ is determined by~\eqref{eq:VsDef}. In
terms of soft parton momenta this gives $T_{m,s}= 
\max_i\{N_i |k_{ix}|/Q\}$, where $k_{ix}$ is the out-of-event-plane
momentum component of emitted parton $i$ and, due to the recoil
kinematics, $N_i$ is $4$ or $2$ respectively according to whether or
not emission 
$i$ is in the same hemisphere as the most energetic hard parton (as
was shown in~\cite{KoutLong} and can easily be determined
numerically as well as analytically).

The analytical result for $\cF^{(\conf)}_{T_m}(R')$ has been computed
in~\cite{KoutLong}.\footnote{Note that in \cite{KoutLong} 
 $\cF$ represented a different quantity.} %
We do not reproduce here its explicit form, since its various
components involve up to as many as five nested integrals, but we plot
in figure~\ref{fig:Kout} the analytical results as a function of $R'$
together with $\cF^{(\conf)}_{T_m}$ obtained using the numerical
procedure.

\EPSFIGURE[ht]{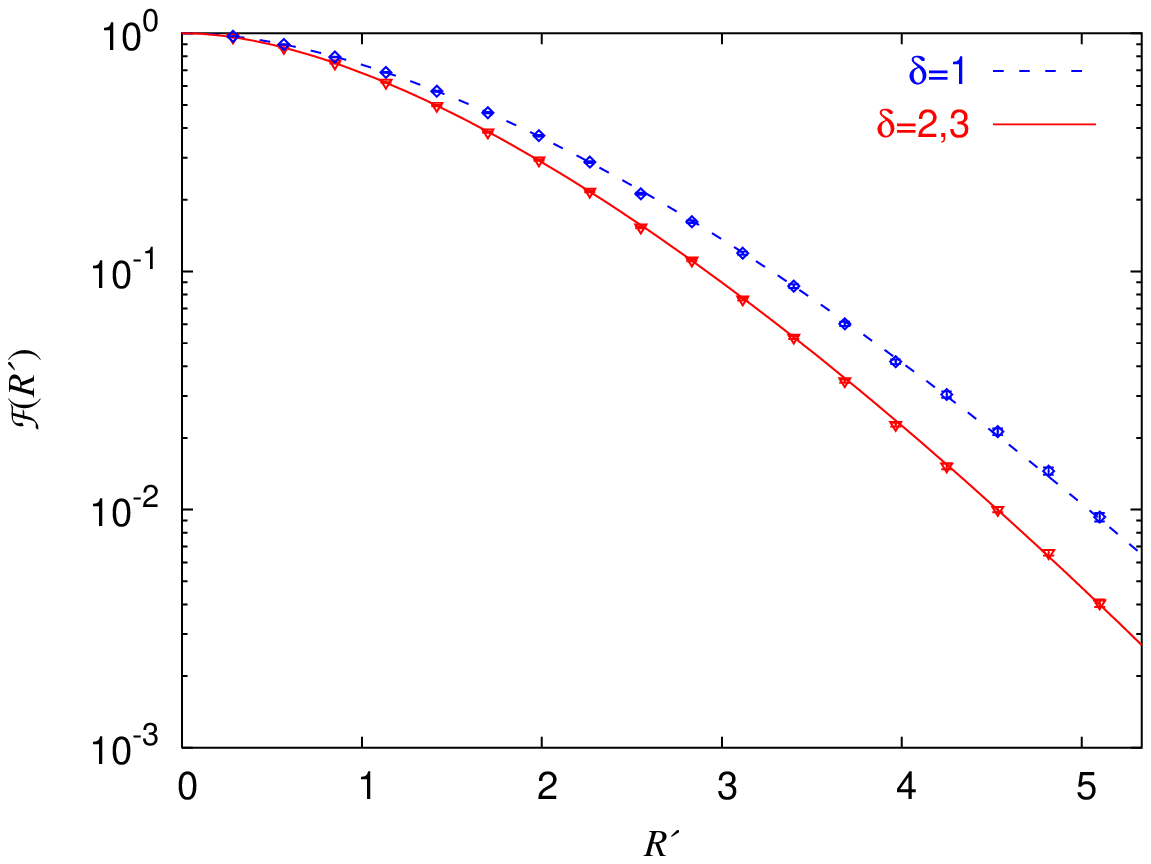,width=0.8\textwidth}{$\cF^{(\conf)}(R')$ for 
the thrust minor distributions in the three different colour configurations. 
The lines indicate the analytical results, while the points are the numerical 
results.\label{fig:Kout}}

\section{New results}
\label{sec:neweg}
We now exploit our method to compute the function $\cF$ for some
observables for which an analytical expression for the resummed PT
distribution has not so far been found: the thrust major, the
oblateness and the Durham three-jet resolution. In the first two cases we
suspect that it may not even be possible to find analytical
expressions.

\subsection{Thrust major}
\label{sec:tmajor}
The thrust major, $T_M$, is defined in \eqref{eq:majordef}. In order to
compute $\cF_{T_M}$ we have as the `simple' observable
\begin{equation}
\label{eq:TM-simple}
T_{M, s}\equiv 2\cdot\max_i  \frac{|\vec{k}_{ti}|}{Q}\>.
\end{equation}
Our numerical procedure gives the function $\cF_{T_M}(R')$ 
shown in figure~\ref{fig:F_TM-O-y3}.  
The resummed $T_M$ distribution is then given by~\eqref{eq:Sigmarelation}, 
where $\Sigma_s(T_M)$, defined generally in \eqref{eq:Sigmas}, is
\EPSFIGURE[ht]{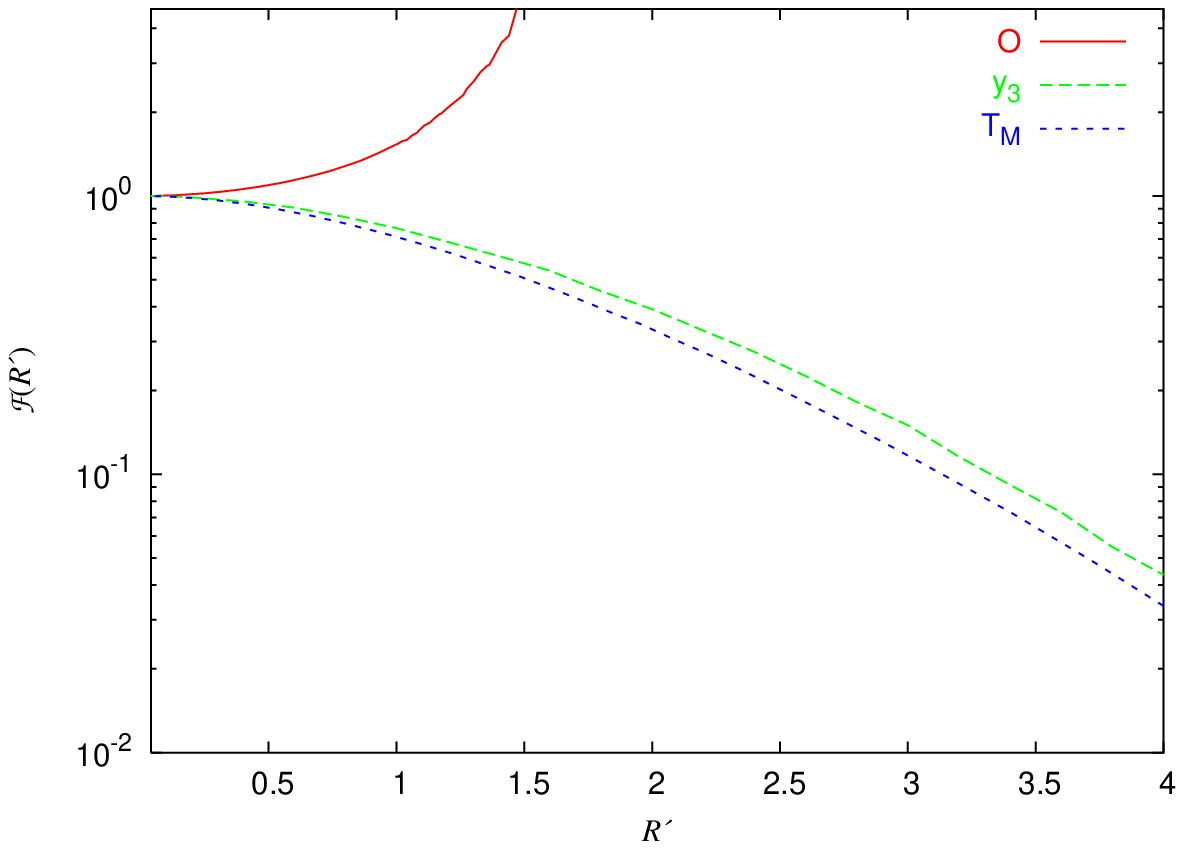,width=0.8\textwidth}{The function $\cF(R')$ for 
the thrust major, the oblateness and the Durham three-jet resolution. 
\label{fig:F_TM-O-y3}}
\begin{equation}
\label{eq:Rad-TM}
\begin{split}
\Sigma_s(T_M)&=e^{-R_s(T_M/2)}=e^{-R_s(T_M)}e^{-R'\ln2}\>,\\
R_s(T_M)&=2C_F\int_{T^2_M Q^2}^{Q^2}\frac{dk_t^2}{k_t^2}\frac{\as(k_t)}{\pi}
\left(\ln\frac{Q}{k_t}-\frac34\right)\>,
\end{split}
\end{equation}
\ie it is expressed in terms of a radiator, $R_s(T_M)$, which is
identical to that for $B_T$ and $B_W$. Its analytical expression to
SL accuracy has been already computed in~\cite{CTW} and is recalled in
Appendix~\ref{app:tm-obl}.

We then match the above resummed result (using the $\log
R$-matching scheme~\cite{CTTW}) with the exact $\cO{\as^2}$ results
computed with EVENT2~\cite{CSDipole}, to give the curve shown in
figure~\ref{fig:TM-Ofull}.
\EPSFIGURE[ht]{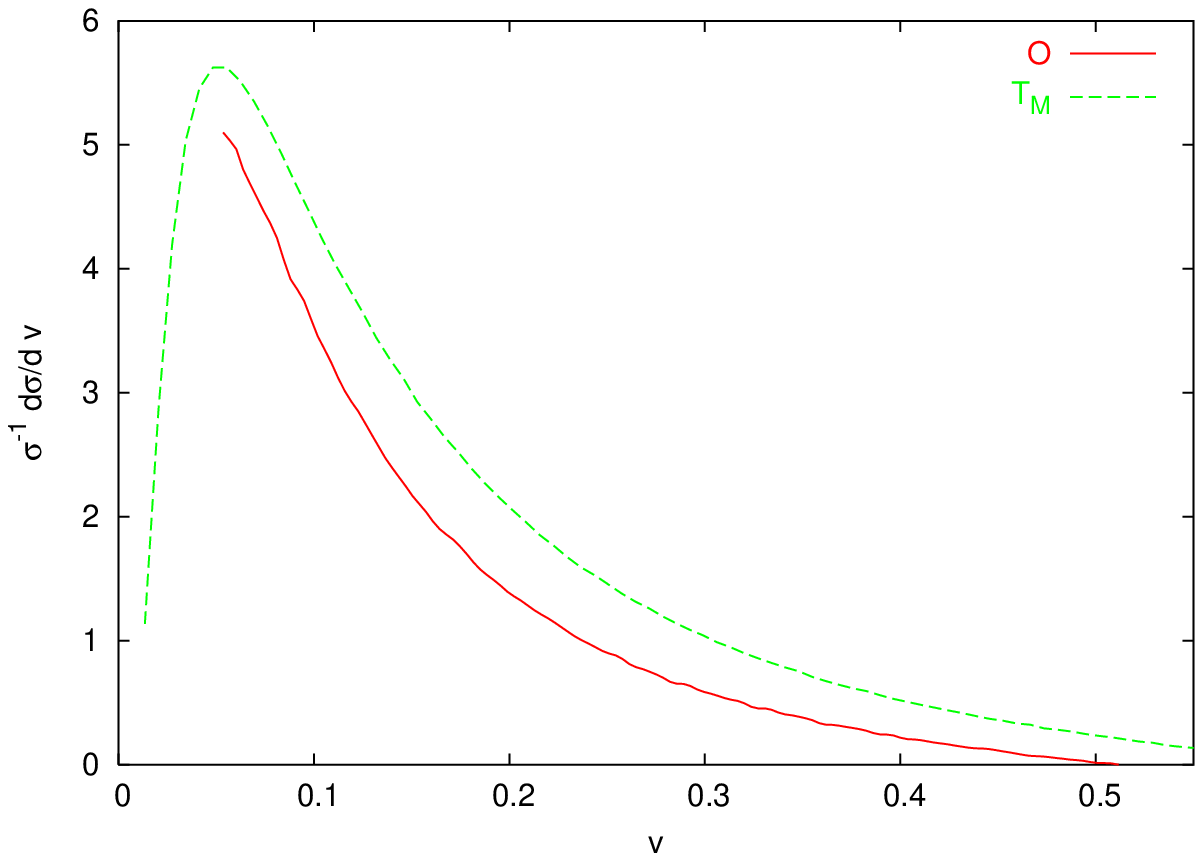,width=0.8\textwidth}{The full SL-resummed 
 distributions for the thrust major and the oblateness including
second order matching ($\log R$-scheme), shown for $\sqrt{s}=M_Z$. 
\label{fig:TM-Ofull}} 

\subsection{Oblateness}

The `simple' observable \eqref{eq:TM-simple} can be also exploited to  
compute $\cF(R')$ for the oblateness, defined as the difference between the
thrust major and the thrust minor (see \eqref{eq:obldef}).  

The function $\cF(R')$ for the oblateness (also shown in
figure~\ref{fig:F_TM-O-y3}) behaves quite differently from that for
the other variables considered so far, since it increases rather than
decreases with increasing $R'$.

There is a simple reason for this difference. For most variables,
adding an extra emission to the ensemble can only increase the value
of the observable, largely because most observables essentially
involve a sum of positive definite quantities.

The oblateness is unusual in that it is a difference of two
quantities. For a single emission the thrust minor is zero and so the
oblateness is equal to the thrust major. But as soon as one considers a
second emission, there exist configurations for which the thrust minor
is almost equal to the major, leading to a value for the oblateness
which is smaller than that of the simple observable.

This is illustrated in figure~\ref{fig:OblDist} which shows the
probability distribution for the ratio of the true variable to the
simple one, for both the thrust major and the oblateness. The thrust
major exhibits a sharp cutoff for $v/v_s < 1$. The oblateness on the
other hand has a roughly power-like tail extending to arbitrarily
small values of $v/v_s$, originating from events with nearly
identical major and minor projections. 
Since eq.~\eqref{eq:FDefn} involves the ratio $(v/v_s)^{-R'}$, if a
sufficient fraction of events has $v < v_s$, then $\cF(R')$ can be
larger than $1$. 

\EPSFIGURE[ht]{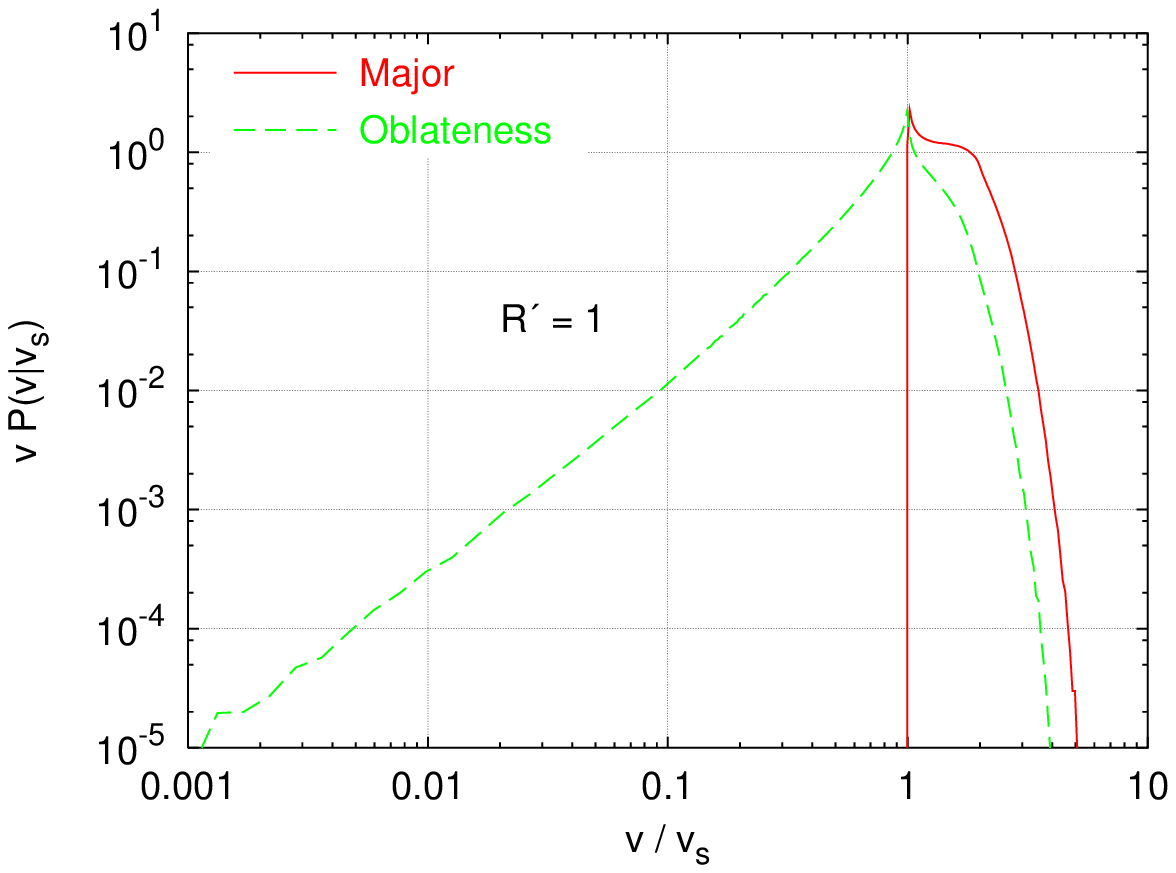,width=0.8\textwidth}{
Distribution of the value of $v/v_s$ for the oblateness and the thrust major.
\label{fig:OblDist}}

However $\cF$ being larger than one is not the only consequence of the
tail at small $v/v_s$: if $R'$ is too large then the factor
$(v/v_s)^{-R'}$ can completely compensate the smallness of $P(v|v_s)$,
and the integral \eqref{eq:DvDsv} then diverges. The start of this
divergence is clearly visible for the oblateness in
figure~\ref{fig:F_TM-O-y3}. We shall refer to the position of the
divergence as $R'_\mathrm{c}$.

In the case of the oblateness, simple considerations based on two-gluon
configurations suggest that $R'_\mathrm{c} = 2$, corresponding to a
tail of $vP(v|v_s)$ proportional to $(v/v_s)^2$. This does not quite
correspond to what is seen in figure~\ref{fig:OblDist}, and it is not
clear whether the actual $vP(v|v_s)$ has a different power for the tail, or
simply some extra logarithmic enhancement.

We emphasise that the divergence in $\cF$ is not a particularity of
our approach. 
Recently for example, such a problem has been extensively discussed in
the context of the jet broadening in DIS \cite{DISBroad}, and it is in
general present for all observables which can have zero value even in
the presence of emissions. Physically, what happens is that for
$R\simeq R'_\mathrm{c}$ there is a transition from the normal
double-log Sudakov suppression to some other behaviour (\eg a
suppression proportional to $v^2$). Such a
transition is beyond representation in terms of a pure SL function,
leading to a breakdown in the hierarchy of leading and subleading logs.
This is seen for example in the fact that NNLL terms are even more
strongly divergent that the NLL ones.  In analytical resummation
approaches, with considerable extra work, it is sometimes possible to
include an appropriate subset of subleading terms so as to bring the
answer back under control over the whole range of the $R'$.

On the other hand as long $R'_\mathrm{c} - R'$ is of order one
(formally $\gg \sqrt{\as}$) it can be shown that the divergence can be
ignored~\cite{DISBroad}. Therefore, so as to be sure of remaining in
the region under control, we suggest that the oblateness distribution
be studied only for $R'(O) <1$. This is illustrated in
figure~\ref{fig:TM-Ofull}, which shows the full resummed matched
distribution for the oblateness (based on an expression analogous to
\eqref{eq:Rad-TM}). The point $R'=1$ is just to the right of the peak
region --- accordingly, it should be possible to compare most, though
not all the data with the resummed predictions.


\subsection{Durham three-jet resolution}
\label{sec:Durham}

Now we wish to calculate the multiple emission single-logs
for the distribution of the three-jet resolution parameter, $y_3$, in the
Durham algorithm (the corresponding integrated distribution is also
referred to as 
the two-jet rate). The leading logs, and some subleading logs (those
necessary for NLL accuracy in the exponentiated answer rather than the
exponent) were calculated in \cite{JetRates}.  The NL logs associated
with the appropriate scheme choice for $\as$ were given in
\cite{DissSchmell}. However those coming from the non-trivial
dependence on multiple emissions have yet to be calculated.

Jet rates differ from most event shapes in that they do not satisfy
item~3 of the conditions for applicability given in
appendix~\ref{sec:applicability}, which was required for a
straightforward elimination of subleading logarithms, using the
event-shape specific method of section~\ref{sec:nosubleading}. This
condition was that $V(\{k_i\})$ should stay invariant under certain
transformations of the $k_i$ which kept the $\{V(k_i)\}$ constant.
It allowed phase-space integrations over rapidity to be carried out
analytically.

Nevertheless we can still use the `general' method for the elimination
of subleading logs, as outlined in section~\ref{sec:nosubleading}. It
requires that we be able to evaluate the observable accurately even
when it takes values much smaller than the smallest machine
representable (double precision) floating point number, so that we can
take the limit $\as \to 0$ with constant $\as L$ (which implies
$L\to\infty$).

It turns out that to succeed in doing this (without recourse to
arbitrary precision arithmetic) we must carry out an analytical
analysis of the observable's dependence on multiple soft and collinear
emissions. One might raise the criticism that this negates
the philosophy of our numerical approach. Such a criticism would be
only partially justified since in a traditional resummation approach
once one has determined this dependence, one still has to understand
what transformations are needed in order to write it in a factorised
form (\cnf the discussion in the introduction), and then one has to
carry out the inverse transformations, both operations involving
considerable work.

\subsubsection{Soft and collinear analysis}

It is useful to start by recalling the definition of the Durham jet
finding algorithm \cite{Durham}, given in terms of a resolution
parameter $y_\mathrm{cut}$ as follows:
\begin{enumerate}
\item For all pairs of (pseudo)particles $i,j$ calculate
  \begin{equation}
    y_{ij} = \frac{2\mathrm{min} (E^2_i, E_j^2) (1 - \cos
      \theta_{ij})}{Q^2}\,.
  \end{equation}
\item If all $y_{ij} > y_\mathrm{cut}$ stop. The number of
  jets is then defined to be equal to the number of (pseudo)particles
  left. 
\item Otherwise \emph{recombine} the pair with the smallest $y_{ij}$
   into a single pseudoparticle 
  (there are various schemes according to whether one wants the
  recombined particle to be massless, see for instance~\cite{JADE}).
\item Go back to step 1.
\end{enumerate}
The three-jet resolution parameter, $y_3$, is the maximum value of
$y_\mathrm{cut}$ that leads to a $3$-jet event.

To understand the soft and collinear limit of our observable, we shall
suppose that we have an event with hard partons $a$ and $b$ and
soft-collinear partons $1\ldots n$. To start with we should examine
the value of the $y_{ij}$ for different kinds of pairings of particles:
\begin{itemize}
\item If $i$ and $a$ are in the same hemisphere $y_{ia} \simeq
  E_i^2\theta_i^2\simeq k_{ti}^2$.
\item If $i$ and $j$ are in different hemispheres then $y_{ij} \simeq
  4E_{<}^2$, where $E_< = \min(E_i,E_j)$.
\item If $i$ and $j$ are in the same hemisphere then $y_{ij} \simeq E^2_<
  |\vec \theta_i - \vec \theta_j|^2$, where we define the angles as
  vectors in the transverse directions.
\end{itemize}
In these relations we have defined ${\vec \theta}_i$ to be the angle between
parton $i$ and its nearest hard parton (the vector indicates the
azimuthal direction), and $k_{ti}$ to be the
relative transverse momentum. The centre of mass energy $Q$ is taken
to be $1$.

To determine the value of $y_3$ for an arbitrary ensemble of soft and
collinear particles it is beneficial to start first with some simple
cases: if we have just a single emission then $y_3=k_t^2$ of that
emission (hence the alternative name --- $k_t$ algorithm).

Now let us consider two emissions, labelled $1$ and $2$. We take
$k_{t1} > k_{t2}$, the non-trivial case being when $k_{t2}$ and
$k_{t1}$ are however of the
same order. We shall assume that the two emissions are
widely separated in rapidity, as is appropriate when considering only
NLL terms.  There are then four relevant configurations, illustrated
in figure~\ref{fig:y3illustr}:
\EPSFIGURE[ht]{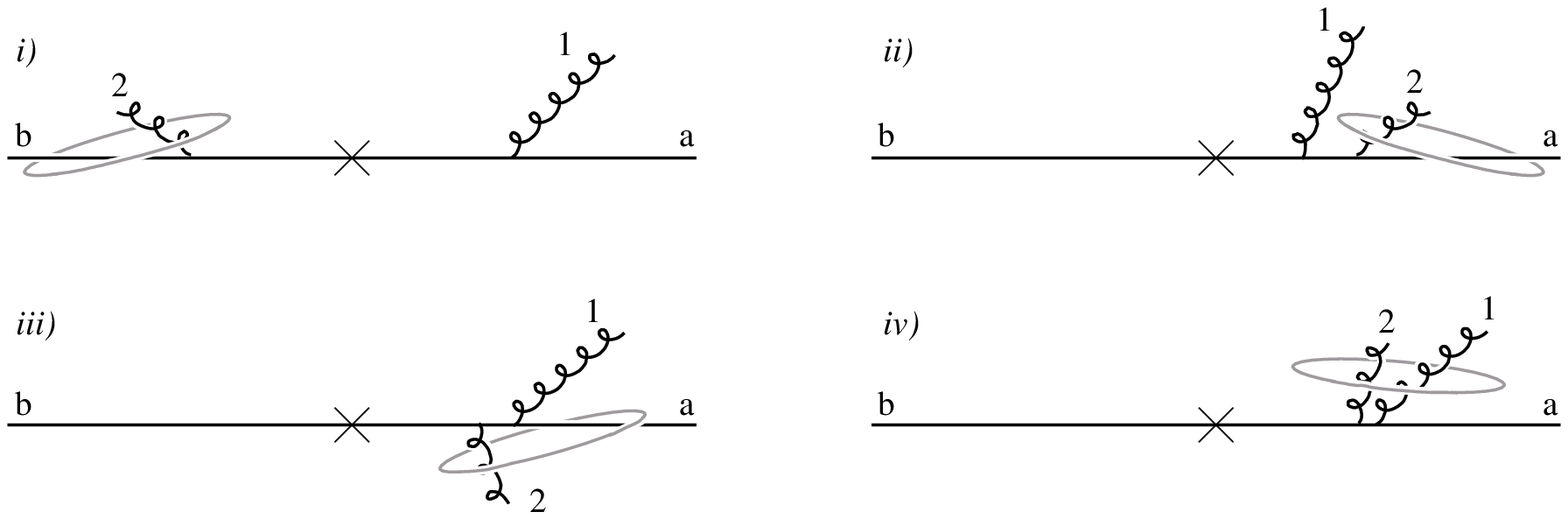,width=\textwidth}{
Representation of the clustering in the Durham
algorithm, for various configurations. Both gluons are soft and
collinear, with $k_{t1} > k_{t2}$ but of the same order. When
they are in the same hemisphere, their angles are assumed to be
strongly ordered (though for clarity only weak ordering is
represented in the figure). The hoop indicates which pair of
particles clusters first.\label{fig:y3illustr}}
\begin{itemize}
\item[\textit{i})] If $1$ and $2$ are in opposite hemispheres (as in
  figure) then $y_{12}=4\min(E_1^2,E_2^2)\gg y_{1a} > y_{2b}$, so
  parton $2$ gets recombined with $b$. Since now the only
  soft-collinear parton left is $1$ we get $y_3 = k_{t1}^2$.
\item[\textit{ii})] If $1$ and $2$ are in the same hemisphere (say the
  hemisphere of 
  $a$) and $\theta_1 \gg \theta_2$ (which implies $E_1 \ll E_2$), then
  we get $y_{12} \simeq y_{1a} = k_{t1}^2$, while $y_{2a} =
  k_{t2}^2$ which is smaller. So once again parton 2 recombines
  with a hard parton and we are left with just parton 1, which then
  leads us to $y_3 = k_{t1}^2$.
\item[\textit{iii})] If $1$ and $2$ are in the same hemisphere (again
  the hemisphere 
  of $a$), but $\theta_1 \ll \theta_2$ (which implies $E_1 \gg E_2$),
  then we have $y_{12} \simeq y_{2a} = k_{t2}^2$, while $y_{1a} =
  k_{t1}^2$ which is larger. So the recombination that will take
  place is either $12$ or $a2$. Which one depends on whether $|\vec
  \theta_2 - \vec \theta_1|^2 > \theta_2^2$. If $\vec \theta_1 \cdot
  \vec \theta_2 < 0$ then the inequality is satisfied, and we have a
  $a2$ recombination, so that just as before $y_3 = k_{t1}^2$. 
\item[\textit{iv})] 
If $1$ and $2$ are in the same hemisphere and $\theta_1 \ll \theta_2$,
as in \textit{iii}), but with 
$\vec \theta_1 \cdot \vec \theta_2 > 0$, then $y_{12} < y_{2a}$ and
(at last a non trivial result!) we recombine partons $1$ and $2$.
This gives us a pseudoparticle with the energy of parton $1$ (recall
$E_1 \gg E_2$) and squared transverse momentum $k_{tp}^2 = |\vec
k_{t1} + \vec k_{t2}|^2$.  The value of $y_3$ is then just set by the
$k_{tp}^2$ of the pseudoparticle. We note that since $\vec \theta_1
\cdot \vec \theta_2 > 0$, $k_{tp}$ is always larger than $k_{t1}$.
\end{itemize}
We can see from this analysis that item~3 of the conditions for
applicability in appendix~\ref{sec:applicability} is violated:
exchanging the rapidities of the two gluons, while keeping their
transverse momenta and azimuths constant (\ie exchanging
configurations \textit{ii}
and \textit{iv}), leads to a change in the value of $y_3$.

The above two-gluon analysis also points the way to a general
algorithm suitable for taking the limit of very small values of $y_3$.
In order to be able to represent all quantities on a computer in
standard double precision we work in terms of rapidities $\eta_i$ and
rescaled
transverse momenta $\kappa_{i} = k_{ti}/\max_j\{k_{tj}\}$. We further
assume (as is appropriate at NLL order) that particles are all widely
separated in rapidity from one another.  The $y_3$ value for the
Durham jet algorithm can then be determined as follows:
\begin{enumerate}
\item Find the index $I$ of the smallest of the $\{\kappa_i \}$. 
\item Considering only the soft partons $j$ in the same hemisphere as
  $I$ and which satisfy $\vec \kappa_I \cdot \vec \kappa_j > 0$, find
  the index $J$ of the one with the smallest positive value of
  $|\eta_j| - |\eta_I|$. If there are no soft
  partons $j$ with both $\vec \kappa_I \cdot \vec \kappa_j > 0$ and
  $|\eta_j| > |\eta_I|$, then let $J=a$ or $b$, according to which 
  is the collinear hard parton.
\item If $J=a$ or $J=b$ then just throw away parton $I$. Otherwise
  recombine $I$ and $J$ as follows: the rapidity $\eta_p$ of the
  pseudoparticle is just set to $\eta_J$ (this involves a mistake on
  $\eta_p$ by an additive amount of order $1$ --- but that is
  subleading), while its transverse momentum is the vector sum, $\vec
  \kappa_p = \vec \kappa_I + \vec \kappa_J$. At NLL order, this
  procedure is
  appropriate for all standard recombination schemes.
\item If only one soft pseudoparticle remains, $P$, then
  $y_3/\max_j\{k_{tj}^2\} = \kappa_P^2$. Otherwise go back to
  step 1.
\end{enumerate}

\subsubsection{Numerical results}

Using the above algorithm for calculating $y_3$ at arbitrarily small
values we can now numerically determine the function
$\cF_{y_3}(R')$. The results are shown in figure~\ref{fig:F_TM-O-y3}
together with those for the thrust major and the oblateness.%
\footnote{We note that the above analysis also makes it quite
  simple to determine $\cF_2$ analytically:
\begin{equation}
         \label{eq:f2-y3}
  \cF_2 = -\frac{1}{4} \int_{-\pi/2}^{\pi/2} 
    \frac{d\phi}{2\pi} \int_0^{k_{t1}^2} \frac{dk_{t2}^2}{k_{t2}^2}
    \log\frac{k_{t1}^2 + k_{t2}^2  + 2k_{t1} k_{t2} \cos
      \phi}{k_{t1}^2} = -\frac{\pi^2}{32}\,,
\end{equation}
where the leading factor of $1/4$ comes because only a quarter of the
time are the particles in the same hemisphere with $\theta_1 \ll
\theta_2$.}

For phenomenology one also needs to know the resummation for the
simple observable related to $y_3$.  Since $y_{3s}$ is just the
(normalised) squared transverse momentum, $y_{3s}=k_t^2/Q^2$, we have
from~\eqref{eq:Sigmas} and~\eqref{eq:Rs} that
\begin{equation}\label{eq:Rsy3}
\begin{split}
\Sigma_s(y_3)=e^{-R_s(y_3)}\,,\qquad 
R_s(y_3)=C_F\int_{y_3 Q^2}^{Q^2}\frac{dk_t^2}{k_t^2}\frac{\as(k_t)}{\pi}
\left(\ln\frac{Q^2}{k_t^2}-\frac32\right)\>.
\end{split}
\end{equation}
This corresponds to what was calculated in \cite{DissSchmell}, which
contains the explicit NLL form for the result of the integration.  The
full NLL resummed prediction for the Durham $y_3$ can then be obtained
from \eqref{eq:Sigmarelation}, with $\cF_{y_3}(R')$ as shown in
figure~\ref{fig:F_TM-O-y3}.

For completeness, in figure~\ref{fig:y3} we show the final
perturbative result for the NLL resummed Durham $y_3$ distribution,
matched to the NLO fixed order results.  
We choose here as a recombination scheme the $E0$ scheme (see for
instance~\cite{JADE}), in which the energy of the recombined
pseudo-particle equals the sum of the energies of the two
(pseudo)-particles, while the three-momentum is rescaled so as to keep
pseudo-particles massless.  Choosing a different recombination scheme
will affect only the large $y_3$ region and subleading logs.
In the same figure we also show the matched resummed curve for the
Cambridge algorithm
\cite{Cambridge}, which is also in the `family' of $k_t$ algorithms
--- however it has a different clustering sequence which leads to the
property that for emissions widely separated in rapidity, the $y_3$
value is simply the maximum of the emitted squared transverse
momenta.%
\footnote{We would like to thank Yu.L.~Dokshitzer
  and B.R.~Webber for bringing this to our attention.} %
As a consequence $\cF(R')=1$, so that the full resummed distribution
is simply given by $\Sigma_s$.

\EPSFIGURE[ht]{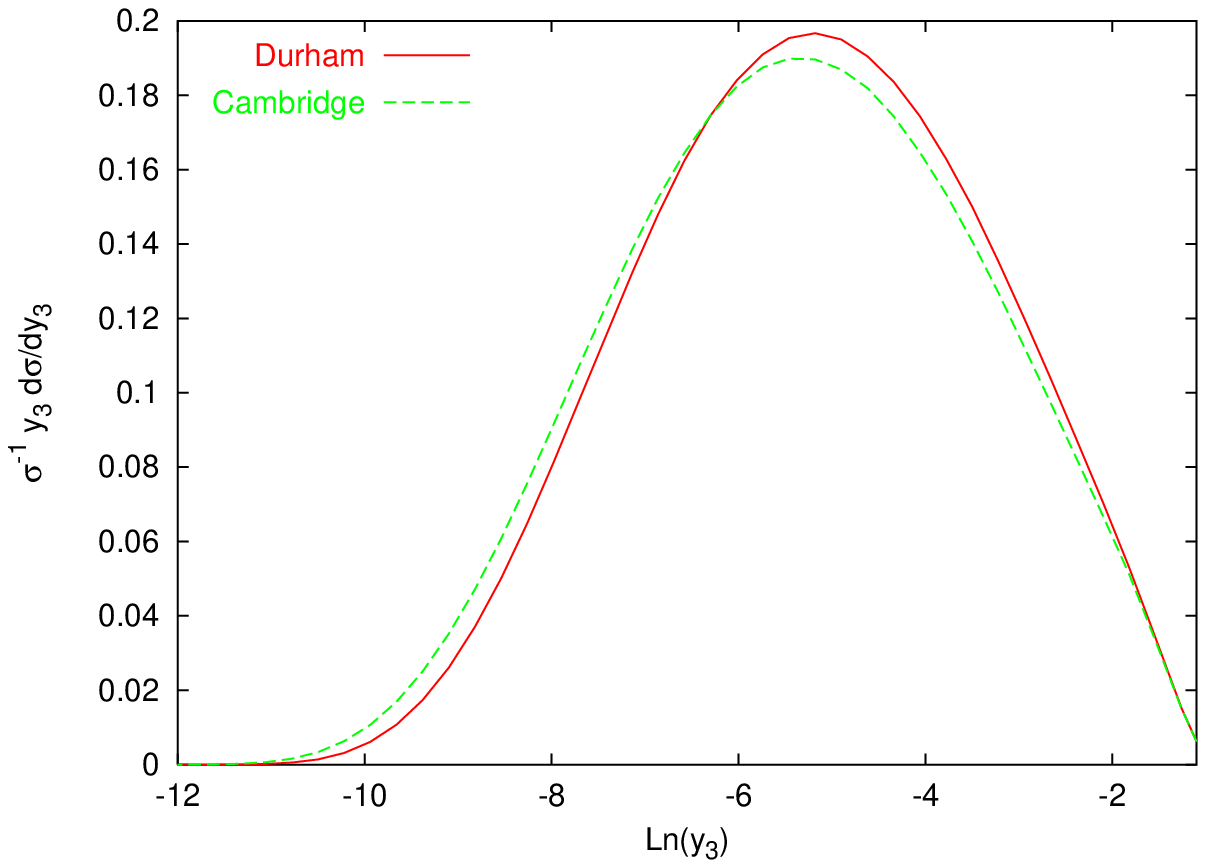,width=0.8\textwidth}{
The full SL-resummed distributions for the Durham and
Cambridge three-jet resolution $y_3$ including second
order matching ($\log R$-scheme).  \label{fig:y3}}

\section{Conclusions}

In this paper we have presented a general method for the numerical
calculation of the only non-trivial class of next-to-leading
logarithms in (global) event shapes. For most event shapes it means
that the calculation of the resummed distribution (in the $n$-jet
limit for an observable which has non-zero values only for ensembles
of $n+1$ or more particles) can be reduced to the following three
tasks:
\begin{itemize}
\item The evaluation of the dependence of the observable on a single soft
  and collinear emission.
\item The calculation of a resummed distribution, $\Sigma_s$, based on that
  single-emission dependence, through eqs.~\eqref{eq:Sigmas} and
  \eqref{eq:Rs} (or their extensions in situations with more than two
  hard `legs', and/or incoming hadrons).
\item The provision of a computer subroutine which calculates the
  value of the observable for an arbitrary set of four-momenta. One
  can then use the algorithm of this paper to determine a single
  logarithmic function $\cF(R')$ which accounts for the observable's
  non-trivial dependence on multiple emissions, and multiplies
  $\Sigma_s$ to give the resummed distribution to full NLL accuracy.
\end{itemize}
The method thus provides both a very simple, but also general way of
obtaining NLL resummed distributions for final-state observables. 

As a cross check, it has been tested against a range of previously
studied observables (including some in processes other than $\ee$,
which for brevity we have not shown here, but which are listed at the
end of appendix~\ref{sec:ObsDef}), and consistently
reproduces the known analytical results.

It has also been applied to observables that up to now had proved
beyond the scope of existing analytical methods, allowing us to
present the first fully NLL resummed distributions for the thrust
major, the oblateness and the three-jet resolution $y_3$ in the Durham
algorithm. For the three-jet resolution some analytical
analysis was required in order to provide a subroutine capable of
giving an accurate evaluation of $y_3$ even in the limit of very
small $y_3$. That analysis also has some intrinsic interest for the
light it casts on the functioning of the Durham jet algorithm.

It should be emphasised that the philosophy of the approach presented
here differs entirely from that of a Monte Carlo event generator (such
as Herwig~\cite{herwig} or Pythia~\cite{pythia}). An event generator may well
reproduce many of the leading and subleading logarithms for event
shape distributions, however even at parton level there will also be
contamination from potentially spurious NNLL and non-perturbative
terms, and it is not currently possible to match to exact
next-to-leading order calculations.

On the other hand the results produced by our algorithm are
indistinguishable from those of fully analytical methods, allowing the
same procedures of truncation at pure LL and NLL terms, and
straightforward matching to fixed leading and NLO calculations. As
such, our method is an essential development should one wish at some
stage to fully automate the calculation of resummed distributions, in
a manner analogous to what has become standard for fixed-order
predictions of non-inclusive observables.


\section*{Acknowledgements}

We wish to thank Pino Marchesini for discussions in the early stages
of this work, Yuri Dokshitzer for helpful suggestions and Stefano
Catani, G\"unther Dissertori and Bryan Webber for useful
conversations on the subject of jet algorithms.

\appendix

\section{Conditions for applicability}
\label{sec:applicability}

Various steps in this article rely on the observable under
consideration being `suitable'. Here we attempt to define what is
meant by this. The first two conditions are required for the
exponentiation of the double logarithms. The third condition is
specific to the second (event-shape specific) method for eliminating
subleading effects 
(discussed in section~\ref{sec:nosubleading}):
\begin{enumerate}
\item If one is resumming from a Born level consisting of $n$ coloured
  hard partons (incoming or outgoing), then for any configuration
  consisting of just $n$ hard partons the observable must have a constant
  value $V_0$ independent of the hard configuration. Furthermore for a
  general configuration of $n+1$ hard partons the observable must have
  a value different from $V_0$. Without loss of generality, one can
  redefine the observable such that $V_0=0$.
  
  So for example, we can consider the thrust in the 2-jet limit, or
  the thrust minor in the 3-jet limit, but not the thrust-minor in the
  2-jet limit, nor the thrust in the 3-jet limit.

\item If we select $m$ soft, collinear momenta (with respect to any
  of the $n$ hard partons) such that $V(k_1)\sim V(k_2)\sim \ldots\sim
  V(k_m)$, then the observable must satisfy the condition
  $V(k_1,\ldots,k_m) \sim V(k_1)$. This is necessary in order for the
  double logarithms to exponentiate, and excludes for example
  observables based on the JADE jet-clustering algorithm
  \cite{JADEproblems}. 
  
\item Given $m$ soft and collinear (SC) partons, the value of the
  observable $V(k_1,\ldots,k_m)$ must remain unchanged (to within
  corrections of relative order of the softness or collinearity) if we
  vary the rapidity and absolute transverse momentum (but not the
  azimuth) of any of the $m$ SC partons subject to the following
  restrictions:
  \begin{itemize}
  \item the $V(k_i)$ are all kept constant;
  \item each SC parton remains collinear to the hard parton to which
    it was originally collinear.
  \end{itemize}
  This condition is necessary if one is to use
  method 2 of section~\ref{sec:nosubleading} for the elimination of
  subleading logs. We note that in general it is satisfied by event
  shapes, but not for example by the jet rate with the Durham
  algorithm.
\end{enumerate}
In addition to the above conditions, we note that for the simple
observable to have a straightforward resummation in terms of
independent emissions (\cnf
eqs.~\eqref{eq:IndepEmsnProb},~\eqref{eq:Sigmas} and \eqref{eq:Rs}) it
is necessary that it be global \cite{1jet}.

\section{Definition of observables}
\label{sec:ObsDef}

For completeness we recall here the definition of all the event shape
variables that we refer to in the main text (except for $y_3$ whose
definition is explicitly needed in section~\ref{sec:Durham} and so
is directly given there).

\begin{itemize}
\item 
{\bf Thrust $T$, Thrust major $T_M$, Thrust minor $T_m$, Oblateness $O$ } 

The thrust is defined as
\begin{equation}
\label{eq:thrustdef}
T=\frac{1}{\sum_i|{\vec p_i}|}
\max_{\vec{n}} \sum_i |\vec{p}_i\cdot \vec{n} |\>,
\end{equation}
where $\vec{p}_i$ are the three-momenta of the outgoing particles.
The thrust axis is the unit vector $\vec{n}_T$ which maximises the sum in 
\eqref{eq:thrustdef}
and gives therefore the direction along which the projection of
momenta is maximal.

The thrust major is then defined similarly, but the maximisation
procedure is restricted to unit vectors perpendicular to $\vec{n}_T$
\begin{equation}
\label{eq:majordef}
T_M=\frac{1}{\sum_i|{\vec p_i}|}
\max_{\vec{n}\cdot \vec{n}_{T}=0}
\sum_i |\vec{p}_i\cdot \vec{n} | \>.
\end{equation}
The vector $\vec{n}_{T_M}$ which maximises the sum
in~\eqref{eq:majordef} defines the thrust major axis. 

Given the thrust and the thrust major axes, the thrust minor is defined as
\begin{equation}
\label{eq:minordef}
T_m=\frac{1}{\sum_i|{\vec p_i}|} 
\sum_i |\vec{p}_i\cdot \vec{n}_{T_m}|\>,
\qquad \vec{n}_{T_m}=\vec{n}_{T}\times\vec{n}_{T_M}\>,
\end{equation}
and the oblateness as 
\begin{equation}
\label{eq:obldef}
O=T_M-T_m\>.
\end{equation}

\item 
{\bf The broadenings:} 

The broadenings measure the transverse size of the jets.  The plane
perpendicular to the thrust axis through the origin divides an event
in two hemispheres, which are usually called left (L) and right (R)
hemisphere. One defines first the right ($B_R$) and left broadening
($B_L$) as
         
\begin{equation}
\label{eq:BR-BL}
B_{R/L}=\frac{1}{2\>\sum_i|{\vec p_i}|} 
\sum_{i\in R/L} |\vec{p}_{ti}|\>,
\end{equation}
with $\vec{p}_{ti}=\vec{p}_i\times\vec{n}_T$. 

One can then define the total ($B_T$) and  wide ($B_W$)
jet broadening as
\begin{equation}
\label{eq:BT-BW}
\begin{split}
B_{T}&=B_{R}+B_{L}\>,\\
B_{W}&=\max\{B_{R},B_{L}\}\>.
\end{split}
\end{equation}

\end{itemize}

As well as having been applied to the observables defined above, our
method has also been successfully tested against known analytical
results for several other variables: the $D$-parameter in 3-jet $\ee$
events \cite{Dpar}, the current-jet broadening (with respect to the
photon axis) in $1+1$-jet DIS events \cite{DISBroad}, the out-of-plane
momentum in $2+1$-jet DIS events \cite{KoutDIS} and the out-of-plane
momentum for $W/Z+\mathrm{jet}$ events in hadron-hadron collisions
\cite{KoutHH}.


\section{Analytical ingredients}
\label{app:matching}

Here we summarise all the analytical ingredients needed for the
calculation of the resummed distributions of the three observables
studied in section~\ref{sec:neweg}. This means NLL expressions for the
simple radiators and derivatives (the formulae are all well known in
the literature), as well as the fixed order expansions.

Computer subroutines which interpolate tabulated numerical values for
the functions $\cF(R')$ are available on request from the authors, as
are example programs for calculating the complete resummed, matched 
distributions.

\subsection{Thrust major and oblateness}
\label{app:tm-obl}
For the thrust major and oblateness, the simple radiator, $R_s$
(defined in \eqref{eq:Rad-TM}) is given to NLL order by $R_s(v) = L
g_{s1}(\as L) + g_{s2}(\as L)$ with
\begin{equation}
  g_{s1}(\as L) = \frac{\CF}{\pi\beta_0\lambda} \left(-2\lambda - \ln
    (1-2\lambda)\right) ,
\end{equation}
and
\begin{multline}
  g_{s2}(\as L) = \frac{3\CF}{2\pi\beta_0} \ln(1-2\lambda)
     + \frac{\CF K}{2\pi^2 \beta_0^2}
            \frac{2\lambda +
              (1-2\lambda)\ln(1-2\lambda)}{1-2\lambda}\\ 
     + \frac{\CF \beta_1}{\pi \beta_0^3}\left( -
       \frac{2\lambda + \ln(1-2\lambda)}{1-2\lambda} -
       \frac12\ln^2(1-2\lambda) 
     \right),
\end{multline}
where $\lambda = \as \beta_0 L$ and $L = \ln 1/v$. The coefficients of
the $\beta$-function are given by
\begin{equation}
  \label{eq:betas}
  \beta_0 = \frac{11\CA - 2\nf}{12\pi},\qquad\quad 
  \beta_1 = \frac{17\CA^2 -5\CA\nf - 3\CF\nf}{24\pi^2}\,,
\end{equation}
and the constant relating the gluon Bremsstrahlung scheme \cite{CMW}
to the $\MSbar$ is
\begin{equation}
K = \CA\left(\frac{67}{18} -
    \frac{\pi^2}{6}\right)
  - \frac{5}{9}\nf\,.
\end{equation}
Also useful is the formula for $R'$ (at NLL equal to $R_s'$):
\begin{equation}
  \label{eq:Rp}
  R'(v) \equiv \frac{d}{dL} (L g_{s1}(\as L)) = 
  \frac{\CF}{\pi \beta_0}  \frac{4\lambda}{1-2\lambda}\,.
\end{equation}

For matching to fixed order calculations, it is necessary to have the
coefficients of the fixed order expansion of the full resummed
distribution. They are given in table~\ref{tab:TM-coef}, using the
notation of~\cite{CTTW}.

\begin{table}[tp]
\begin{center}\begin{tabular}{|ccl|}\hline
 & &\\
$G_{12}$ &=& $-4C_F$ \\
 & &\\
$G_{11}$ &=& $C_F\left(6-8\ln2\right)$ \\
 & &\\
$C_1$ &=& $C_F \left(6\ln 2 -4 \ln^2 2 + \pi^2-\frac{17}{2}\right)$\\
 & &\\
$G_{23}$ &=& $-C_F\left(\frac{88}{9}C_A-\frac{32}{9}T_R n_f\right)$ \\
 & &\\
$G_{22}$ &=& $ C_F\left(64 \cF_2 C_F 
+\left(\frac23 \pi^2-\frac{35}{9}-\frac{88}{3}\ln 2\right) C_A
+\left(\frac49+\frac{32}{3}\ln2\right)T_R n_f\right)$ \\
 & &\\
\hline \end{tabular}
\caption{Logarithmic coefficients $G_{nm}$ and the constant $C_1$ for
thrust major and oblateness; for $T_M$ one has $\cF_2=-0.46851\pm
0.00017$, for the oblateness $\cF_2= 0.37799\pm  0.00029$.}
\label{tab:TM-coef}
\vspace*{-0.25cm}
\end{center}
\end{table}

Tables~\ref{tab:TM:g21-c2}a and \ref{tab:TM:g21-c2}b show the two
subleading coefficients $G_{21}$ and $C_2$ for the thrust major and
the oblateness respectively. They are obtained from the fixed order
integrated distribution (calculated with EVENT2 \cite{CSDipole}) after
subtracting the leading and next-to-leading logs. The final errors
shown are obtained by adding in quadrature the mean error and the
maximal discrepancy between the mean coefficient and the coefficient
obtained in a single fit. The errors are to be considered as
indicative only, because of the difficulty in estimating the
systematic uncertainty associated with the choice of fit range.

\begin{table}[tp]
\begin{center}
\begin{tabular}{|c||c|c|}
\multicolumn{3}{c}{a) Thrust Major}
\\\hline
Fit-range & $G_{21}$ & $C_2$ \\ \hline
$[-9;-5]$ & $-20.3\pm 2.5$ &  $  154.0\pm 14.8 $ \\
$[-8;-5]$ & $-20.7\pm 3.3$ &  $  156.2\pm 19.3 $ \\
$[-9;-6]$ & $-21.0\pm 2.9$ &  $ 160.4\pm 19.2  $ \\
\hline
          & $-20.6\pm 2.9$ &  $  156.9\pm  18.1 $\\
\hline
\end{tabular}\qquad
\begin{tabular}{|c||c|c|}
\multicolumn{3}{c}{b) Oblateness}
\\\hline
Fit-range & $G_{21}$ & $C_2$ \\ \hline
$[-8;-4]$ & $75.3\pm 1.0$ &  $166.1 \pm 4.6   $ \\
$[-7;-4]$ & $75.6\pm 1.1$ &  $164.9 \pm 5.3 $ \\
$[-8;-5]$ & $70.8\pm 2.9$ &  $191.6 \pm 16.5 $ \\
\hline
        & $73.9\pm 3.5$ &  $174.2 \pm 19.5 $\\
\hline
\end{tabular}
\caption{Fits of $G_{21}$ and $C_2$ for the thrust major and the oblateness.} 
\label{tab:TM:g21-c2}
\end{center}
\end{table}

\subsection{Three-jet resolution}

For the Durham (and Cambridge) three jet resolution parameter, the
simple radiator, $R_s$, defined in \eqref{eq:Rsy3}, is given to NLL
order by $R_s(v) = L g_{s1}(\as L) + g_{s2}(\as L)$ with
\begin{equation}
  g_{s1}(\lambda) = \frac{\CF}{\pi\beta_0\lambda} \left
    [-\lambda-\ln{(1-\lambda)}\right]\,,
\end{equation}
\begin{multline}
g_{s2}({\lambda})=  \frac{3\CF}{2\pi \beta_0} \ln(1-\lambda)
  +\frac{K\CF [\lambda + (1-\lambda)\ln(1-\lambda)]}
  {2\pi^2\beta_0^2(1-\lambda)} 
  \\\nonumber  +
  \frac{\CF\beta_1}{\pi\beta_0^3} \left[ -\frac{\lambda + \ln
      (1-\lambda)}{1-\lambda} - \frac12 \ln^2{(1-\lambda)} \right],
\end{multline}
and, as before, $\lambda=\as\beta_0 L$. 
$R'$ is given by 
\begin{equation}
  R' = \frac{\CF}{\pi \beta_0} \frac{\lambda}{1-\lambda}\,.
\end{equation}
The coefficients of the fixed order expansion of the resummation are
given in table~\ref{tab:y3-coef}.

\begin{table}[tp]
\begin{center}\begin{tabular}{|ccl|}\hline
 & &\\
$G_{12}$ &=& $-C_F$ \\
 & &\\
$G_{11}$ &=& $3C_F$ \\
 & &\\
$C_1$ &=& $C_F \left(-6\ln 2 + \frac{\pi^2}{6}-\frac{5}{2}\right)$\\
 & &\\
$G_{23}$ &=& $-C_F\left(\frac{11}{9}C_A-\frac{4}{9} T_R n_f\right)$ \\
 & &\\
$G_{22}$ &=& $ C_F\left( 
4 \cF_2
C_F
+\left(\frac16 \pi^2-\frac{35}{36}\right) C_A
+\frac{T_R n_f}{9}\right)$ \\
 & &\\
\hline \end{tabular}
\caption{Logarithmic coefficients $G_{nm}$ and the constant $C_1$ for
the three-jet resolution parameter. In the Durham algorithm $\cF_2 =
-\pi^2/32$, while in the Cambridge algorithm $\cF_2 = 0$.}
\label{tab:y3-coef}
\vspace*{-0.25cm}
\end{center}
\end{table}

From fixed order results we also extract the subleading coefficients,
shown in tables~\ref{tab:y3:g21-c2dur}a and \ref{tab:y3:g21-c2dur}b
for the Durham and Cambridge algorithms respectively. We follow the
same fit procedure as for the thrust major and the oblateness.  
Note that these coefficients are obtained in the $E0$-scheme (see for
example~\cite{JADE}), and that choosing a different recombination
scheme will affect their value.

\begin{table}[tp]
\begin{center}
\begin{tabular}{|c||c|c|}
\multicolumn{3}{c}{a) Durham}
\\\hline
Fit-range & $G_{21}$ & $C_2$ \\ \hline
$[-13;-8]$ & $-7.4\pm 0.1$ &  $  19.3 \pm 1.0 $ \\
$[-13;-9]$ & $-7.0\pm 0.1$ &  $  15.8 \pm 1.3  $ \\
$[-12;-8]$ & $-7.4\pm  0.1$ &  $ 19.4 \pm 1.1 $ \\
\hline
          & $-7.2\pm 0.3$ &  $  18.2\pm  2.6 $\\
\hline
\end{tabular}\qquad
\begin{tabular}{|c||c|c|}
\multicolumn{3}{c}{b) Cambridge}
\\\hline
Fit-range & $G_{21}$ & $C_2$ \\ \hline
$[-13;-8]$ & $1.9\pm 0.1$ &  $  -30.5 \pm 1.2 $ \\
$[-13;-9]$ & $2.3\pm 0.2$ &  $  -34.1 \pm 1.6 $ \\
$[-12;-8]$ & $1.9\pm  0.2$ &  $ -30.3 \pm 1.4 $ \\
\hline
          & $2.1\pm 0.3$ &  $  -31.6\pm 2.9  $\\
\hline
\end{tabular}
\caption{The coefficients $G_{21}$ and $C_2$ for the three-jet
  resolution in the Durham and Cambridge algorithms.}
\label{tab:y3:g21-c2dur}
\end{center}
\end{table}


\end{document}